\documentclass[11pt]{amsart}
\usepackage{graphicx}
\usepackage{amssymb}
\usepackage{epstopdf}
\usepackage{MnSymbol}
\usepackage{subfigure}

\newtheorem{lemma}{Lemma}[section] 
\newtheorem{propos}[lemma]{Proposition}
\newtheorem{prop}[lemma]{Proposition}
\newtheorem{example}[lemma]{Example}

\newtheorem{cor}[lemma]{Corollary}

\newtheorem{defn}[lemma]{Definition}
\newcommand{\ev}{\mathrm{ev}}

\newcommand{\extd}{\mathrm{d}}
\newcommand{\cc}{\mathrm{c}}

\newcommand{\tens}{\mathop{\otimes}}
\newcommand{\la}{{\triangleright}}
\newcommand{\ra}{{\triangleleft}}

\newcommand{\id}{\mathrm{id}}

\newcommand{\<}{\langle}
\renewcommand{\>}{\rangle}

\title{The Majid-Ruegg model and the Planck scales}
\author{E.J.\ Beggs \& R.M.\ Makki}
\address{College of Science, Swansea University, Swansea, SA2 8PP U.K.}

\date{}                                           % Activate to display a given date or no date

\begin{document}

%\section{}
%\subsection{}

\begin{abstract}
A novel differential calculus with central inner product is introduced for $\kappa$-Minkowski space.  
The `bad' behaviour of this differential calculus is discussed with reference to symplectic quantisation and $A$-infinity algebras. Using this calculus in the Schr\"odinger equation gives two values 
which can be compared with the Planck mass and length. This comparison gives an approximate numerical value for the deformation parameter in $\kappa$-Minkowski space. 
We present numerical evidence that there is a potentially observable variation of propagation speed in the Klein-Gordon equation.
The modified equations of electrodynamics (without a spinor field) are derived from noncommutative covariant derivatives. We note that these equations suggest that the speed of light is independent of frequency, in contrast to the KG results (with the caveat that \textit{zero current} is not the same as \textit{in vacuum}).
We end with some philosophical comments on measurement related to quantum theory and gravity (not necessarily quantum gravity) and noncommutative geometry. 
\end{abstract}

\maketitle

\section{Introduction}

The Majid-Ruegg or $\kappa$-Minkowski (kappa being the deformation parameter) space time is a well studied model of noncommutative geometry related to physical theories. It is related to a deformation of the Poincar\'e group as a Hopf algebra, called the $\kappa$-Poincar\'e algebra, which coacts on $\kappa$-Minkowski space. The description of the algebra as a bicrossproduct can be found in \cite{majid-ruegg}. 
A discussion of $\kappa$-Minkowski space as a topological algebra is given in \cite{DuSi}.
We can give the $\kappa$-Poincar\'e algebra a bicovariant differential calculus (of one dimension more than the classical case) or a left covariant differential calculus (of the classical dimension), after the manner of \cite{woron87}, and these differential calculi descend to $\kappa$-Minkowski space. This has been done in several papers on theoretical physics, including \cite{majid-ruegg,MercatiKappa}.  The Klein-Gordon equation in the extra dimension case is discussed in \cite{majidNewtonianGrav}. 
Field theory and gauge theory on the space can be found in \cite{DiJoMo} and \cite{MercatiKappa},
and \cite{PrGhPa} considers dynamics of charged particles.
A non-relativistic particle model related to $\kappa$-Minkowski is discussed in \cite{ghoshPalKappa}.
(Another noncommutative version of Maxwell's theory is found in \cite{DoPe}.)
In \cite{DiJo} it is noted that their twisting approach to coupling fermions in 
$U(1$) gauge theory is not fully consistent with the $\kappa$-Poincar\'e symmetry.

Of particular relevance to this paper are results on the varying speed of light, and the related observational test of noncommutativity, proposed in \cite{Am-CaMaj}, and the study of electrodynamics in \cite{HarikMaxwells,HaJuMeElectro}. 

In this paper we shall begin with $\kappa$-Minkowski space, or more precisely a noncommutative algebra which is the corresponding deformation of functions on classical Minkowski space. If we look at its commutation relations in isolation, ignoring the $\kappa$-Poincar\'e algebra, we see one reason why it is such an interesting model of physics: It is rotationally invariant, but not Poincar\'e invariant. This matches cosmological observations of the universe, which is rotationally invariant (in the large, though this is now being strained by statistical studies of galaxy distributions), but where the cosmic microwave background provides a reasonably well defined rest frame. 

However, we shall diverge from previous literature in giving a different differential structure, unrelated to the Hopf algebra differential calculi. This is done by imposing the condition that functions commute with the Minkowski metric to first order in the deformation parameter $\lambda$, which gives a unique deformed calculus, to first order. 
This commutativity is a natural and extremely useful property, as it allows us to make a well defined inner product in a fiberwise tensor product of fields, and we shall use it in looking at noncommutative electromagnetism. 
To explain the commutativity requirement quite bluntly, without it the Minkowski metric is not usable in its traditional sense, it is not possible to simply introduce the metric into a tensor expression in a well defined fashion. 
In \cite{BeMaStarGravInd} an effort was made to find a deformation of the Minkowski metric which was central for the usual calculus derived from the $\kappa$-Poincar\'e algebra, but it was found that this is not possible in 3+1 dimensions, though it was possible in 1+1. 
To restate the situation, this commutativity is not true for the differential structure derived from the $\kappa$-Poincar\'e algebra action, but there is no physical reason why this algebra must necessarily act. 
As far as we know, our choice of calculus has never been considered before in the literature.

To begin with the bad news, this is a very badly behaved calculus. It is quite easy to see that it is not associative at $O(\lambda^2)$. This can be remedied by introducing an explicit associator, but in turn that associator does not satisfy the pentagon identity at $O(\lambda^4)$, so we need to introduce higher order corrections. It is quite possible that the corrections required to patch up the usual laws of algebra never stop, so why do we not throw the calculus away? One reason is that the higher order corrections come with larger powers of $\lambda$, and $\lambda$ is so small that, even without the associator, the lack of associativity would likely not be observable. Another reason is that in mathematics and theoretical physics $A_\infty$-algebras \cite{BKellerAinfty,KonSoi09}, structures which have this sort of infinite number of corrections, are now widely studied. A third reason is that it is expected that gravity messes up the laws of calculus, or stated rather less colourfully, in \cite{hawkRig} and \cite{BeMaSemiClass} it is shown that the differential calculus of an algebra given by symplectic quantisation is not associative (the algebra of functions by itself, of course, has an associative deformation) if the curvature (in cases, a measure of gravitational field) is not zero. Currently, there is no neat way around this problem in symplectic quantisation, and repairing the nonassociativity would likely require a messy order by order correction. As we shall see, our calculus displays some gravitational effects. Also remember that nonassociativity is intrinsically no barrier to geometry or physical theories  \cite{BHMnonass,BeMaNonAssTwist,majidNonAss}

Some good news is that we can do very explicit calculations, and get classical results plus small corrections, but this can also be done for the standard calculus.  We have a well behaved Riemannian inner product, which is not true for the other calculus. In fact, the bad behaviour of the Riemannian metric in the usual calculus is $O(\lambda)$, whereas even if we do not apply an associator, the bad behaviour in the calculus we propose is only $O(\lambda^2)$.
However the most striking feature of this calculus is the nature of the physical results it gives. 
The Schr\"odinger equation comes with a correction, of the form of a critical or limiting mass $\hbar/(\cc^2\,|\lambda|)$. By some speculation, we could equate this with the Planck mass $m_P=\sqrt{\hbar\,c/G}$, and use this to find a numerical value of $|\lambda|$. The behaviour of the `renormalised' mass in the Schr\"odinger equation can also be examined for small values of mass, giving a speculative link to the Planck length and a value of $|\lambda|$. 
If (and it is a big IF) this model makes physical sense, it is because the noncommutativity is causing similar effects to those proposed for the origins of the Planck length and Planck mass. 

The massless Klein-Gordon equation has a possible correction to its propagation speed.  We give numerical evidence that this effect is real. Given the proposed value of $\lambda$, we see how measurable this effect is. This should be compared to the discussion with the usual calculus in \cite{Am-CaMaj}. 

We discuss $U_1$ gauge theory, or electrodynamics, in this calculus. It turns out that we have small nonlinear corrections to the classical equations (with no spinor field), but there is no obvious variation in the speed of light. (There is a caveat on \textit{in vacuo} not quite being the same thing as \textit{zero current}.) We conclude with a short discussion on introducing noncommutative geometry into physics, by looking at measurement problems related to geometry and quantum theory.

The authors would like to thank S.\ Majid for discussions and assistance.

\section{Preliminaries on differential graded algebras}
For space time we take coordinates $ x_0=t,x_{1}=x,x_2=y,x_3=z $. The Minkowski metric is given by $c^2\, dt\otimes dt-dx\otimes dx-dy\otimes dy-dz\otimes dz $, where $c$ is the velocity of light. This is written as $ \sum\eta_{ab}\,dx_{a}\otimes dx_{b} $.

The Majid-Ruegg or $ \kappa $-deformed space time is given by making this noncommutative.
We have the following commutator relations, where $ \lambda $ is an imaginary number,  using  the convention that $ i,j,k $ take values $ 1,2,3 $ only:
\begin{equation}  \label{vuksavc}
[x_{i},x_{j}] = 0\ ,\quad
[x_{i},x_0] = \lambda x_{i}\ .
\end{equation}
Applying the differential $\extd$ to (\ref{vuksavc}) gives the following relations:
\begin{equation}  \label{vuksavc2}
[\extd x_{i},x_{j}] = [\extd x_{j},x_{i}] \ ,\quad
[\extd x_{i},x_0] -[\extd x_{0},x_i]= \lambda \,\extd x_{i}\ .
\end{equation}
The algebra is normally given a differential calculus derived from the theory of quantum groups. However, we do something completely different, we deform the differential calculus according to the principle that the metric is central, i.e., it commutes with all functions, so the expression in (\ref{cbfhjwkl}) vanishes:
\begin{eqnarray}\label{cbfhjwkl}
\quad\left[\sum\eta_{ab}\,dx_{a}\otimes dx_{b}\,,\,x_{c}\right] \,=\, \sum\eta_{ab}([dx_a,x_c]\otimes dx_{b}+dx_{a}\otimes [dx_{b},dx_{c}])\ .
\end{eqnarray}
To fit with (\ref{vuksavc2}) we shall assume the following form for the commutators of functions and forms:
\begin{equation} \label{vbcdksv}
[dx_{a},x_{b}]=\Lambda^{s}_{ab}\,\,dx_{s}\ .
\end{equation}
Now (\ref{vuksavc2}) gives constraints on the numbers $\Lambda^{s}_{ab}$,
\begin{eqnarray}
\Lambda^{s}_{ij}=\Lambda^{s}_{ji}\ ,\quad \Lambda^{s}_{i0}-\Lambda^{s}_{0i} &=& \left\lbrace{\begin{array}{c}\lambda\quad s=i \\ 0\quad s\neq i\end{array}}\right.\ .
\end{eqnarray}
There is a unique solution $\Lambda^{s}_{ab}\in\mathbb{C}$ to this problem, given by
 all commutation relations $ [dx_{a},x_{b}]=0 $, except for $ (i=1,2,3) $
\begin{eqnarray}
\left[dx_{i},x_{i}\right] \,=\, -\lambda\,c^2\,\,dx_{0} \ ,\quad
\left[dx_{0},x_{i}\right] \,=\, -\lambda\,\,dx_{i}\ .
\end{eqnarray}
Applying $\extd$ to (\ref{vbcdksv}) gives the usual relations for the higher order calculus:
\begin{eqnarray}
\extd x_a\wedge\extd x_b+\extd x_b\wedge\extd x_a\,=\,0\ .
\end{eqnarray}
Every associative algebra obeys the Jacobi identity for commutators $ [a,b]=ab-ba $,
\begin{equation}  \label{cbdhjkshcvk}
0=J(a,b,c)=[[a,b],c]+[[b,c],a]+[[c,a],b]
\end{equation}
However if we calculate the Jacobi identity, we find the following:
\begin{equation}
[x_{0},[dx_{0},x_{i}]]+[dx_{0},[x_{i},x_{0}]]+[x_{i},[x_{0},dx_{0}]] = -\lambda^{2}\,\, dx_{i}\ .
\end{equation}
Now the reader may be concerned that to obtain (\ref{cbfhjwkl}) we used the associative law in rearranging the brackets. In fact we only need the algebra to be associative to $O(\lambda)$ here, and the nonassociativity only appears at $O(\lambda^2)$.

\section{Repairing associativity}
Similarly to the situation in \cite{hawkRig,BeMaSemiClass}, we have a deformed algebra (functions on a noncommutative space) that is associative, but a differential calculus that is not.  Similarly to \cite{hawkRig,BeMaSemiClass}, where the lack of asasociativity is caused by curvature, we shall find that possibly gravitational effects lurk behind the nonassociativity. 
Similarly to \cite{hawkRig,BeMaSemiClass}, there is not any obvious easy way to repair the lack of associativity. 
In this section we shall begin to repair the lack of associativity, but it is not obvious that we will not need an infinite number of repairs. However needing an infinite number of `repairs' is not an unheard of situation -- see the theory of $A$-infinity algebras \cite{BKellerAinfty,KonSoi09}. 

One good thing from a physical point of view is that $\lambda$ is a small parameter, so the first order corrections are small in normal physical situations. The $O(\lambda^2)$ lack of associativity is smaller, and by constructing an associator $\Phi$ we push the anomalies into yet higher powers of $\lambda$. The reader who is happy with associativity just working to first order in $\lambda$ is welcome to skip this section, and who is to say that we should expect associativity at all? 

An associator is a machine for altering the order of bracketing on tensor products, $\Phi_{A,B,C}:(A\otimes B)\otimes C\to A\otimes (B\otimes C)$. A product $\mu$ is associative with respect to the associator if, where id is the identity,
\begin{eqnarray}
\mu(\mu(a\tens b)\tens c)\,=\,\mu(\id\tens\mu)\Phi((a\tens b)\tens c)\ .
\end{eqnarray}
 In ordinary vector spaces, associative algebras etc.\ $\Phi$ is trivial, all it does is to change the positions of the brackets. However we need a more complicated form, for elements $a,b,c$ of $A,B,C$:
\begin{equation} \label{cbksyvyu}
\Phi\big((a\otimes b)\otimes c\big)=a\otimes (b\otimes c)+\lambda^{2}\phi\big(a,b,c\big)
\end{equation}
The condition for the product $\mu(a\tens b)=a\,b$ to be associative with respect to the associator in (\ref{cbksyvyu}) is
\begin{eqnarray}
(ab)c-a(bc) &=& \lambda^{2}\mu(id\otimes \mu)\phi(a, b,c)\ .
\end{eqnarray}
We can rewrite $J(a,b,c)$ of (\ref{cbdhjkshcvk}) as
\begin{eqnarray*}
J(a,b,c) &=& \big((ab)c-a(bc)\big)-\big((ba)c-b(ac)\big)+\big((bc)a-b(ca)\big)\\
&&+ \big((ca)b-c(ab)\big)-\big((ac)b-a(cb)\big)-\big((cb)a-c(ba)\big)\ .
\end{eqnarray*}
If we use $\mu^2\phi=\mu(id\otimes \mu)\phi$ for short, we can write
\begin{eqnarray}
J(a,b,c) &=& \lambda^{2}\left(\begin{array}{c}\mu^2\phi(a,b,c)-\mu^2\phi(b,a,c)+\mu^2\phi(b,c,a)
\\
+\mu^2\phi(c,a,b)-\mu^2\phi(a,c,b)-\mu^2\phi(c,b,a)\end{array}\right) 
\end{eqnarray}

Now $ \Phi $ should satisfy the pentagon identity \cite{StasheffPentagon} (essentially this means that rebracketing depends only on the initial and final brackets, not on the intermediate steps), meaning that the following diagram Fig.\ 1 commutes:

\unitlength 0.6 mm
\begin{picture}(110,90)(-30,0)
\linethickness{0.3mm}
\multiput(80,75)(0.14,-0.12){208}{\line(1,0){0.14}}
\put(110,50){\vector(4,-3){0.12}}
\linethickness{0.3mm}
\multiput(85,15)(0.12,0.15){167}{\line(0,1){0.15}}
\put(85,15){\vector(-3,-4){0.12}}
\linethickness{0.3mm}
\multiput(5,50)(0.12,0.12){208}{\line(1,0){0.12}}
\put(5,50){\vector(-1,-1){0.12}}
\linethickness{0.3mm}

\multiput(4,42)(0.12,-0.24){125}{\line(0,-1){0.24}}
\put(19,12){\vector(1,-2){0.12}}

\linethickness{0.3mm}
\put(50,8){\line(1,0){20}}
\put(50,8){\vector(-1,0){0.12}}
\put(55,80){\makebox(0,0)[cc]{$((A\tens B)\tens C)\tens D$}}

\put(115,45){\makebox(0,0)[cc]{$(A\tens (B\tens C))\tens D$}}

\put(115,65){\makebox(0,0)[cc]{$\Phi_{A,B,C}\tens \id$}}

\put(105,55){\makebox(0,0)[cc]{}}

\put(97,8){\makebox(0,0)[cc]{$A\tens ((B\tens C)\tens D)$}}

\put(112,25){\makebox(0,0)[cc]{$\Phi_{A,B\tens C,D}$}}

\put(60,15){\makebox(0,0)[cc]{$\id\tens\Phi_{B,C,D}$}}

\put(22,8){\makebox(0,0)[cc]{$A\tens (B\tens (C\tens D))$}}

\put(2,45){\makebox(0,0)[cc]{$(A\tens B)\tens (C\tens D)$}}

\put(105,80){\makebox(0,0)[cc]{}}

\put(115,65){\makebox(0,0)[cc]{$\Phi_{A,B,C}\tens \id$}}

\put(0,65){\makebox(0,0)[cc]{$\Phi_{A\tens B,C,D}$}}

\put(-5,24){\makebox(0,0)[cc]{$\Phi_{A,B,C\tens D}$}}

\put(150,8){\makebox(0,0)[cc]{\textbf{Fig. 1}}}

\end{picture}

Using the formula (\ref{cbksyvyu}), to order $ \lambda^2 $ this becomes:
\begin{equation}
\phi(a,b,c)\otimes d+\phi(a,b\otimes c,d)+a\otimes\phi(b,c,d)=\phi(a\otimes b,c,d)+\phi(a,b,c\otimes d)\ .
\end{equation}
If we use
\begin{eqnarray*}
t(b)=\left\lbrace{\begin{array}{c}0\quad b=1,2,3 \\ 1\quad b=0\quad\quad\end{array}}\right. \quad ,\quad s(a)=\left\lbrace{\begin{array}{c}0\quad a=1,2,3 \\ 1\quad a=0\quad\quad\end{array}}\right.
\end{eqnarray*}
we can give the following form for $J$:
\begin{equation}
J(dx_{c},x_{a},x_{b}) = \lambda^{2}\left(\begin{array}{c}dx_{a}\,\delta_{bc}\big(t(c)\,s(a)-\mathrm{c}^2\,s(c)\big) \\ -dx_{b}\,\delta_{ac}\big(t(c)\,s(b)-\mathrm{c}^2\,s(c)\big)\end{array}\right)\ .
\end{equation}
If we suppose that the objects that the associator is applied to can have partial derivative applied to them (such as functions, forms and vector fields), then 
we can try writing (summing over $ a,b $)  
\begin{eqnarray}
\phi(x,y,z) &=& \alpha_{ab}(x)\otimes\big(\dfrac{\partial y}{\partial x_{a}}\otimes \dfrac{\partial z}{\partial x_{b}}\big) + \dfrac{\partial x}{\partial x_{a}}\otimes\big(\beta_{ab}(y)\otimes \dfrac{\partial z}{\partial x_{b}}\big)\nonumber
\\
&&+\,\dfrac{\partial x}{\partial x_{a}}\otimes\big(\dfrac{\partial y}{\partial x_{b}}\otimes \gamma_{ab}(z)\big)\ .
\end{eqnarray}
Just isolating the $\alpha$ term of this equation in the pentagon identity (we add the $\beta$ and $\gamma$ terms, which are similar) we get:
\begin{eqnarray*}
 \lambda^2\big(\alpha_{ab}(w)\otimes x+w\otimes \alpha_{ab}(x)-\alpha_{ab}(w\otimes x)\big)\otimes\dfrac{\partial y}{\partial x_a}\otimes\dfrac{\partial z}{\partial x_b}
\end{eqnarray*}
If $ \alpha_{ab}(w\otimes x)=\alpha_{ab}(w)\otimes x+w\otimes \alpha_{ab}(x) $, and similarly for $\beta$ and $\gamma$, we get a solution to the pentagon identity, i.e., if $ \alpha,\beta,\gamma $ acts on a tensor product as an element of a lie algebra. 

We make a choice $\alpha=\gamma=0$ and
\begin{eqnarray}
\phantom{cvdisyic} 2\,\beta_{f,e}(dx_{c})\,=\,\extd x_{e}\, \delta_{f,c}(t(c)s(e)-s(c))-\extd x_{f}\, \delta_{e,c}(t(c)s(f)-s(c))\ .
\end{eqnarray}
This action on 1-forms can be extended to tensor products by the formula 
$ \beta_{ab}(w\otimes x)=\beta_{ab}(w)\otimes x+w\otimes \beta_{ab}(x) $. Note that $\beta$ is purely a pointwise linear map in the cotangent space - it sends functions to zero. This means that the associativity of the product of functions is not disturbed. Note that $ \beta_{12} $ (and $ \beta_{ij} $) are infinitesimal rotations (i.e.\ in the lie algebra of the rotation group). However, $ \beta_{01} $ is not an infinitesimal Lorentz boost.

There are conditions to check about conjugate modules and star maps in the presence of a nontrivial associator
(see \cite{barcats}), but it turns out that these are satisfied to $O(\lambda^2)$ by the usual formulae, if we take
$\overline{\beta_{ab}(\xi)}=\beta_{ab}(\overline{\xi})$. 

The $ \lambda^4 $ terms appearing when applying $ \Phi $ clockwise around Fig.\ 1, starting with $((u\tens v)\tens w)\tens x$ are:
\begin{eqnarray*}
&& \lambda^4 \dfrac{\partial^2 u}{\partial x_a\partial x_c}\otimes\beta_{cd}(\beta_{ab}(v)\otimes\dfrac{\partial x}{\partial x_b})\otimes \dfrac{\partial x}{\partial x_d}+\lambda^4 \dfrac{\partial u}{\partial x_a}\otimes\dfrac{\beta_{ab}(v)}{\partial x_c}\otimes\beta_{ab}(\dfrac{\partial w}{\partial x_b})\otimes\dfrac{\partial x}{\partial x_d}\\
&&+ \lambda^4 \dfrac{\partial u}{\partial x_a}\otimes(\dfrac{\partial x}{\partial x_b}\otimes\beta_{cd})\beta_{ab}(v\otimes w)\otimes\dfrac{\partial^2 x}{\partial x_d\partial x_b}
\end{eqnarray*}
while the term we get anticlockwise is:
\begin{eqnarray*}
\lambda^4 (\dfrac{\partial }{\partial x_c}\otimes\beta_{cd})\dfrac{\partial (u\otimes v)}{\partial x_a}\otimes\dfrac{\partial }{\partial x_d}(\beta_{ab}(w)\otimes\dfrac{\partial x}{\partial x_b})\ ,
\end{eqnarray*}
and a brief check shows that they do not cancel, i.e.\ the Pentagon identity is not satisfied at $ O(\lambda^4) $.

\section{Normal ordering}
It will be convenient to identify elements of the noncommutative algebra $A$ with ordinary functions on space time.
As the spatial coordinates $\{x_1,x_2,x_3\}$ commute, we can combine them into a single function $\theta(x_1,x_2,x_3)$. By convention write the function of time first, so a general element of the algebra would be a sum of $\phi(x_0)\,\theta(x_1,x_2,x_3)$. The commutator of the functions of the spatial and temporal variables is, to $O(\lambda)$
\begin{eqnarray}  \label{vcdksuvff}
[\theta(x_1,x_2,x_3),\phi(x_0)]\ =\ \lambda\,\frac{\partial \phi}{\partial x_0}  \sum_i x_i\, 
\frac{\partial \theta}{\partial x_i} \ .
\end{eqnarray}
 If $\phi$ is a function of time and $\theta$ a function of space, then $:\!\phi\,\theta\!:$ is an ordinary function on space time whose value is $:\!\phi\,\theta\!:\!(x_0,x_1,x_2,x_3)=\phi(x_0)\,\theta(x_1,x_2,x_3)$. 
The cost of this identification is that the product is no longer the ordinary product of functions on space time: for $\psi,\chi\in A$, (\ref{vcdksuvff}) gives to $O(\lambda)$
\begin{eqnarray} \label{vcdjkbvhjk}
:\!\psi\,\chi\!: \,=\, :\!\psi\!:\,:\!\chi\!: +\lambda\sum_i x_i\,\frac{\partial :\!\psi\!:}{\partial x_i}\, \frac{\partial :\!\chi\!:}{\partial x_0}\ .
\end{eqnarray}
There is also a correction to the star map:
\begin{eqnarray} \label{bvhislzggi}
:\!\psi^*\!: \,=\, :\!\psi\!:^* +\lambda\sum_i x_i\,\frac{\partial^2 :\!\psi\!:^*}{\partial x_i\, \partial x_0}\ .
\end{eqnarray}
We extend normal order to forms $:\!\xi\!:$, given by (from left to right) function of time, function of space, then any $\extd x_a$ (if present). 
Then for a 1-form $\xi=\sum\xi_a\,\extd x_a$,
\begin{eqnarray}  \label{bchdosuoiuv}
:\!\xi^*\!: \,=\,:\!\xi\!:^* + \lambda\,\sum_i \Big(x_i\frac{\partial^2 :\!\xi\!:^*}{\partial x_0\, \partial x_i}
-\frac{\partial :\!\xi_0\!:^*}{\partial x_i}\,\extd x_i-\cc^2\,\frac{\partial :\!\xi_i\!:^*}{\partial x_i}\,\extd x_0\Big)\ .
\end{eqnarray}
For electrodynamics we shall be talking about antiHermitian 1-forms for the noncommutative algebra, which are not very easy to work with. On the other hand, classical antiHermitian forms are simply forms with imaginary valued components. It will be very convenient to use a 1-1 correspondence between noncommutative antiHermitian 1-forms and 2-forms
and classical antiHermitian forms:

\begin{propos} \label{bvcahjksjhc}
There is a 1-1 correspondence between classical antiHermitian 1-forms $\alpha$ and noncommutative
antiHermitian 1-forms $\xi$ given by setting $\alpha=\alpha_a\,\extd x_a$ to be the antiHermitian part of $:\!\xi\!:$ one way, and the other
\begin{eqnarray*} 
:\!\xi\!:\,=\, \alpha + \frac\lambda 2\sum_i \Big(x_i\frac{\partial^2 \alpha}{\partial x_0\, \partial x_i}
-\frac{\partial \alpha_0}{\partial x_i}\,\extd x_i-\cc^2\,\frac{\partial \alpha_i}{\partial x_i}\,\extd x_0\Big)\ .
\end{eqnarray*}
Similarly there is a 1-1 correspondence between classical antiHermitian 2-forms $\beta$ and noncommutative
antiHermitian 1-forms $\omega$ given by setting $\beta=\beta_{ab}\,\extd x_a\wedge\extd x_b$ to be the antiHermitian part of $:\!\omega\!:$ one way, and the other
\begin{eqnarray*}
:\!\omega\!: &=& \beta + \frac\lambda 2\sum_i \Big(x_i\frac{\partial^2 \beta}{\partial x_0\, \partial x_i}
-\frac{\partial \beta_{a0}}{\partial x_i}\,\extd x_a\wedge\extd x_i     -     \frac{\partial \beta_{0b}}{\partial x_i}\,\extd x_i\wedge\extd x_b \cr
&& -\ \cc^2\,\frac{\partial \beta_{ai}}{\partial x_i}\,\extd x_a\wedge\extd x_0 -  \cc^2\,\frac{\partial \beta_{ib}}{\partial x_i}\,\extd x_0\wedge \extd x_b\Big)\ .
\end{eqnarray*}
\end{propos}

\section{Vector fields and partial derivatives} \label{cbvhdksukuv}
As usual, the vector fields are the dual of the 1-forms. We use $\partial_a=\frac{\partial}{\partial x_a}$ as a basis of the vector fields, with the evaluation map $\ev(\partial_a\tens\extd x_b)=\delta_{a,b}$. This means that we act by vector fields written on the left. 
The complication in calculating the derivative of a function is that the 1-forms do not commute with functions. We get, for $\theta(x_1,x_2,x_3)$ to $O(\lambda)$,
\begin{eqnarray} \label{ncjipdsa}
\extd\theta(x_1,x_2,x_3)\ =\ \sum_i\extd x_i\,\frac{\partial \theta}{\partial x_i} +\frac{\lambda\,c^2}{2}\, \extd x_0\sum_i \frac{\partial^2 \theta}{\partial x_i^2}\ .
\end{eqnarray}
Then we get, for a general element with $\phi(x_0)$,
\begin{eqnarray} \label{ncjipdsasadc}
\extd(\phi\,\theta)\,=\, \sum_i\extd x_i\,\phi\,\frac{\partial \theta}{\partial x_i} +\extd x_0\,\frac{\partial \phi}{\partial x_0}\,\theta +\frac{\lambda\,c^2}{2}\, \extd x_0\sum_i \phi\,\frac{\partial^2 \theta}{\partial x_i^2}\ .
\end{eqnarray}
Now we can calculate the noncommutative derivatives of $\phi\,\theta$ along the coordinate directions:
\begin{eqnarray}  \label{cbdhks}
\qquad\ev(\partial_0\tens \extd(\phi\,\theta)) \,=\, \frac{\partial \phi}{\partial x_0}\,\theta +\frac{\lambda\,c^2}{2}\, \sum_i \phi\,\frac{\partial^2 \theta}{\partial x_i^2}\ ,\ \  \ev(\partial_i\tens \extd(\phi\,\theta)) \,=\,\phi\,\frac{\partial \theta}{\partial x_i} 
\end{eqnarray}
Using the normal order we can rewrite (\ref{cbdhks}) for $\psi\in A$ as
\begin{eqnarray} \label{bchdlsdvblv}
:\frac{\partial \psi}{\partial x_0}: \,=\, \frac{\partial :\!\psi\!:}{\partial x_0} +\frac{\lambda\,c^2}{2}\, \sum_i \frac{\partial^2 :\!\psi\!:}{\partial x_i^2}\ ,\quad :\frac{\partial \psi}{\partial x_i}: \,=\, \frac{\partial :\!\psi\!:}{\partial x_i}\ .
\end{eqnarray}

\section{The Schr\"odinger equation} \label{cvgfkucjy}
Classically the Schr\"odinger equation is
\begin{eqnarray} \label{bvcyaisugvi}
\mathrm{i}\,\hbar\,\frac{\partial\Psi}{\partial t}\,=\, -\,\frac{\hbar^2}{2\,m}\,\sum_i\, \frac{\partial^2\Psi}{\partial x_i^2}+\Psi\,V\ ,
\end{eqnarray}
where $\Psi$ is the complex valued wave function of a particle of mass $m$, and $V$ is a real potential, depending only on $x_1,x_2,x_3$. So what is the form of the Schr\"odinger equation in our noncommutative space-time with $\Psi\in A$? One guess is that we replace the partial derivatives in (\ref{bvcyaisugvi}) by the noncommutative partial derivatives (\ref{bchdlsdvblv}) and take normal order to get
\begin{eqnarray} \label{aisugvi}
\mathrm{i}\,\hbar\,\frac{\partial:\!\Psi\!:}{\partial t}+ \frac{\mathrm{i}\,\hbar\,\lambda\,\cc^2}{2}\, \sum_i \frac{\partial^2 :\!\Psi\!:}{\partial x_i^2}\,=\, -\,\frac{\hbar^2}{2\,m}\,\sum_i\, \frac{\partial^2 :\!\Psi\!:}{\partial x_i^2}+:\!\Psi\!:\,V\ .
\end{eqnarray}
This can be rewritten as
\begin{eqnarray} \label{aisugvvsvi}
\mathrm{i}\,\hbar\,\frac{\partial:\!\Psi\!:}{\partial t}\,=\, -\,\Big(\frac{\hbar^2}{2\,m}+\frac{\mathrm{i}\,\hbar\,\lambda\,c^2}{2}\Big)\,\sum_i\, \frac{\partial^2:\!\Psi\!:}{\partial x_i^2}+:\!\Psi\!:\,V\ .
\end{eqnarray}
Remembering that $\lambda$ is imaginary, this is the classical Schr\"odinger equation again, but with a modified real mass, 
\begin{eqnarray}  \label{nbclool}
m'\,=\, m\,\Big(1+\frac{\mathrm{i}\,m\,\lambda\,\cc^2}{\hbar}\Big)^{-1}\ .
\end{eqnarray}

\section{The Klein-Gordon equation} \label{bhucdosv}
Classically the Klein-Gordon equation \cite{ryderbook} is
\begin{eqnarray} \label{bhclslhv}
\frac{1}{\cc^2}\,\frac{\partial^2 \psi}{\partial t^2}-\frac{\partial^2 \psi}{\partial x^2}-\frac{\partial^2 \psi}{\partial y^2}-\frac{\partial^2 \psi}{\partial z^2}\,=\,-\,\frac{m^2\,\cc^2}{\hbar^2}\,\psi\ .
\end{eqnarray}
If we use the noncommutative partial derivatives (\ref{bchdlsdvblv}), and normal order for $\psi\in A$, this becomes to $O(\lambda)$
\begin{eqnarray} \label{bhclslhv5}
\quad\frac{1}{\cc^2}\,\frac{\partial^2 :\!\psi\!:}{\partial t^2}
+\lambda\sum_i \frac{\partial^3 :\!\psi\!:}{\partial t\,\partial x_i^2}
-\frac{\partial^2 :\!\psi\!:}{\partial x^2}-\frac{\partial^2 :\!\psi\!:}{\partial y^2}-\frac{\partial^2 :\!\psi\!:}{\partial z^2}\,=\,
-\,\frac{m^2\,\cc^2}{\hbar^2}\,:\!\psi\!:\ .
\end{eqnarray}
If we substitute a plane wave $:\!\psi\!:=\exp(\mathrm{i}(\omega\,t+\alpha_1\,x_1+\alpha_2\,x_2+\alpha_3\,x_3))$ into this we get, where $|\alpha|^2=\sum\alpha_i^2$,
\begin{eqnarray}  \label{vcgycyuc}
\omega^2-m^2\,\cc^4/\hbar^2=\cc^2\,|\alpha|^2(1-\mathrm{i}\,\lambda\,\omega)\ .
\end{eqnarray}
We shall assume that the rest mass $m$ of the particle is reasonably small, i.e.\ $m\,\cc^2/\hbar\ll 1/|\lambda|$.
Then as $|\alpha|^2\ge 0$, we see that $\hbar\,|\omega|\ge m\,\cc^2$. 
The energy for the Klein-Gordon equation is given by the operator $E=\mathrm{i}\,\hbar\,\frac{\partial}{\partial t}$
and the momentum is $p_i=-\,\mathrm{i}\,\hbar\,\frac{\partial}{\partial x_i}$,
 which, given the noncommutative partial derivative, assigns the following energy and momentum to the plane wave above:
\begin{eqnarray}
E\,=\, -\,\hbar(\omega+\mathrm{i}\,\lambda\,\cc^2\,|\alpha|^2/2)\ ,\quad p_i\,=\,\hbar\,\alpha_i\ .
\end{eqnarray}
Substituting (\ref{vcgycyuc}) into the equation for the energy gives
\begin{eqnarray}
E\,=\, -\,\hbar\,\frac{\omega-\mathrm{i}\,\lambda\,\omega^2/2-\mathrm{i}\,\lambda\,m^2\,\cc^4/(2\,\hbar^2)
}{1-\mathrm{i}\,\lambda\,\omega}
\end{eqnarray}
Now for $|\omega|$ small, we see that $\omega$ is a negative real number to ensure that the energy is positive. Remembering our discussion on the Schr\"odinger equation in Section~\ref{cvgfkucjy}, and its conclusion that $\mathrm{i}\,\lambda>0$, we can study the behaviour of the energy as $\omega$ tends to $-\infty$. For small negative $\omega$ the energy is approximately $E\cong -\hbar\,\omega$, whereas for large negative $\omega$, 
$E\cong -\hbar\,\omega/2$. The transition between these formulae takes place in the region of $\omega=-1/|\lambda|$.

\section{The Planck mass, the Planck length and the Planck time} \label{ncioab}
{\textbf{The Planck mass.}}
The formula (\ref{nbclool}) for the effective mass $m'$ in the Schr\"odinger equation
shows that in the case $\mathrm{i}\,\lambda>0$ there is a maximum effective mass (for $m>0$) of
$m_{\mathrm{crit}}=\hbar/(\cc^2\,|\lambda|)$. For the case $\mathrm{i}\,\lambda<0$ there is a singularity in the effective mass for $m>0$. 
Setting this value $m_{\mathrm{crit}}$ equal to the Planck mass $m_P=\sqrt{\hbar\,\cc/G}$ gives 
$|\lambda|=\cc^{-5/2}\sqrt{\hbar\,G}$, which is the Planck time.

{\textbf{The Planck length.}}
For small values of mass $m,n$ in the Schr\"odinger equation, we have, to lowest order in the masses
\begin{eqnarray}
m'+n'-(m+n)'\,=\,2\,\mathrm{i}\,\lambda\,m\,n\,c^2/\hbar\ .
\end{eqnarray}
This has parallels with the idea of renormalised or dressed mass from quantum field theory -- the idea being that corrections within the theory alter the value of the initial parameters. 
This comparison between the effective masses \textit{suggests} that if we were to combine particles of (small) effective mass $m'$ and $n'$ to make a single particle, we would get a binding energy of $2\,\mathrm{i}\,\lambda\,m\,n\,c^4/\hbar$.
That, for stability, we would like a positive binding energy gives $\mathrm{i}\,\lambda>0$. 
 The product of masses here suggests that we should compare this with the gravitational binding energy (Newtonian potential energy) $G\,m\,n/r$, where the two masses would be brought to within distance $r$ of each other. Thus, utterly hypothetically, we have identified the process of merging two `point' particles into a single `point' particle with bringing the two particles into a distance $r$ of each other. That `merging distance' is $r=G\,\hbar/(2\,|\lambda|\,\cc^4)$, and setting this to be the Planck length $\ell_P=\sqrt{\hbar\,G/\cc^3}$ gives $|\lambda|=\cc^{-5/2}\sqrt{\hbar\,G}/2$, half the Planck time.

{\textbf{The value of $\lambda$.}}
We have, approximately, $G\cong 6.67 \times 10^{-11}\, m^3\,K\!g^{-1}\,s^{-2}$, 
$\hbar\cong 1.05 \times 10^{-34}\, m^2 K\!g\, s^{-1}$ and $\cc\cong 3.00\times 10^8\,m\,s^{-1}$. 
This gives the Planck time $\cc^{-5/2}\sqrt{\hbar\,G}\cong 5.37\times 10^{-44}\,s$. Our estimates of $|\lambda|$ are based on the Planck scales. The idea of the Planck length is not precisely definable. It depends on the idea of classical length breaking down at a certain scale, but yet is calculated based on the idea that classical length works up to that limit. Such a method is likely to introduce some uncertainty into the numerical value. 
The two estimates we have discussed based on the modified mass are to some extent independent (one estimate using large and the other small masses). To assign an uncertain numerical value, we take $|\lambda|$ to be the Planck time $\cc^{-5/2}\sqrt{\hbar\,G}\cong 5.37\times 10^{-44}\,s$. This value is expected, e.g.\ see \cite{Am-CaMaj}.
But what is the sign of $\mathrm{i}\,\lambda$? Seemingly it may be random, as we have no way of standardising $\mathrm{i}=\sqrt{-1}$. However this is not quite true, in writing the Schr\"odinger equation we may be choosing a preferred value of $\mathrm{i}=\sqrt{-1}$. Now our discussion about the sign of $\mathrm{i}\,\lambda$ suggests that it should be positive. The choice is fairly obvious, we go for the limiting mass rather than a singularity, and we go for the positive binding energy rather than the negative one. We write
\begin{eqnarray}
\lambda\,\cong\, -\,5\!\cdot\!37\,\mathrm{i}\, \times 10^{-44}\,s\ .
\end{eqnarray}

\section{Observing noncommutativity}

{\textbf{The varying velocity of waves.}} 
To maximise the chance of observing an effect of noncommutativity, we should take an effect which is classically zero, but becomes nonzero for $\lambda\neq 0$. The speed of plane waves in the massless Klein-Gordon equation is $\cc$, but if we examine (\ref{vcgycyuc}) we see that the phase speed for $m=0$ is $\cc\,\sqrt{1-\mathrm{i}\,\lambda\,\omega}$. 
We make a caveat: It is not obvious that this is the correct speed to take. For example, the massive Klein-Gordon equation has a phase velocity for plane waves which is greater than $c$, but no information can be transmitted at that speed \cite{ShoreOptics,berrySuper}. 

The phase speed being $v$ is simply saying that the plane wave is of the form $f(x-v\,t)$. The more useful idea is the group velocity which, given a sudden happening, is the speed at which the effects propagate. The theory here is not obvious to the authors - unlike the massive KG equation, the superluminal velocity does not tend to $c$ as $|\omega|\to\infty$. The noncommutative equation has a third order differential, and we know of no analysis of such equations. Before continuing to the well worn estimates of how observable this effect is, we present numerical evidence that the effect really exists.
 
 \medskip
 
{\textbf{Numerical models of the massless 1+1D kappa KG equation.}} 
A simple model to examine is a sawtooth initial condition, with value zero on an interval to make the propagation speed more obvious. We choose parameters $\cc=1$, $\mathrm{i}\,\lambda=1$ and $\hbar= 1$, which is of course completely unphysical, but the aim is to illustrate whether the speed of light really can be exceeded. 
Fig.\ 2 shows a graph showing the real part of the usual KG equation on the left and kappa KG on the right, for the same one period of a sawtooth initial condition. In all the graphs, $x$ is on the left axis and $t$ on the right.
\begin{center}
  \includegraphics[scale=.45]{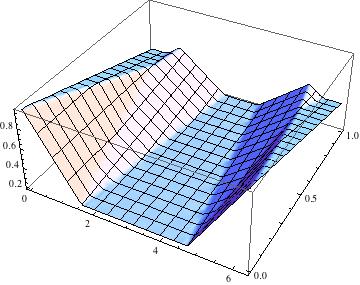}
   \includegraphics[scale=.45]{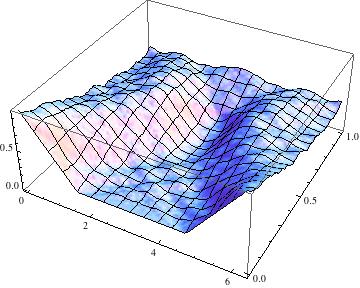}
  
\textbf{Fig.\ 2. Standard (left) and kappa (right) KG equation}
  \end{center}
This was done by using the following Mathematica commands, where $f[x, n]$ is the Fourier series of the sawtooth function, for the initial condition at  $t=0$.
{\scriptsize{\begin{verbatim}
Plot3D[Re[Pi/8 + Sum [f[x, n]/2 Exp[-I n t], {n, 1, 30}]], {x, 0, 2 Pi},
 {t, 0, 1}, PlotPoints -> {40, 40}] 
Plot3D[Re[Pi/8 + Sum [f[x, n]/2 Exp[-I n/2 (n + Sqrt[4 + n^2]) t], 
{n, 1, 30}]], {x, 0,2 Pi}, {t, 0, 1}, PlotPoints -> {40, 40}]
\end{verbatim}
}}

As can be seen, the classical case exhibits an obvious finite propagation speed, with a straight line wavefront. The kappa case has nonzero solutions propagating seemingly much faster. An analysis of the truncation errors involved in the sums suggests that this is a real effect.
However the reader may prefer a non-periodic example. The complication is that this requires Fourier transforms. Fig.\ 3 shows a graph showing the real part of the usual KG equation on the left and kappa KG on the right, for the same bell curve initial condition $1/(1+x^8)$.

\begin{center}
  \includegraphics[scale=.4]{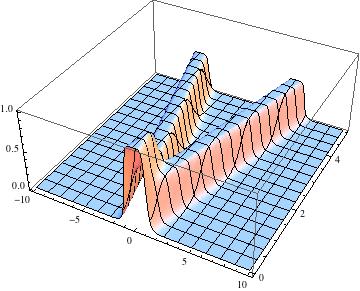}   \includegraphics[scale=.4]{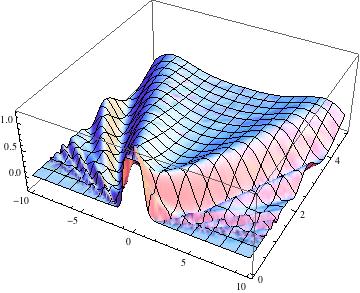}
  
\textbf{Fig.\ 3. Standard (left) and kappa (right) KG equation}
  \end{center}
  
In Fig.\ 3 the classical constant velocity propagation can be seen on the left. However on the right we see separation into different frequency components. The Mathematica code for the classical case is:
  
{\scriptsize{\begin{verbatim}
Plot3D[Re[ 1/4 (-1)^(1/8) 1/ 2 (NIntegrate[(E^((-1)^(1/8) a) 
+ (-1)^(1/4)  E^((-1)^(3/8) a) -  I E^(-(-1)^(5/8) a) 
- (-1)^(3/4) E^(-(-1)^(7/8) a)) Exp[ I a t] Exp[-I  a x], 
{a, -Infinity,  0}] + NIntegrate[(E^(-(-1)^(1/8) a) + (-1)^(1/4)
E^(-(-1)^(3/8) a) -  I E^((-1)^(5/8) a) - (-1)^(3/4)
E^((-1)^( 7/8) a)) Exp[-I a t] Exp[-I  a x], {a, 0, Infinity}])],
 {x, -10, 10}, {t, 0, 5},  PlotPoints -> {40, 40}]
\end{verbatim}
}}

Note that we used NIntegrate, Mathematica's numerical integration method, for the inverse Fourier transform. In fact, we could have worked out the result exactly, but chose this as the following kappa case cannot (as far as we know) be calculated exactly, and we wanted to use the same methods. 

{\scriptsize{\begin{verbatim}
Plot3D[Re[1/4 (-1)^(1/8) 1/2 (NIntegrate[(E^((-1)^(1/8) a) +
 (-1)^(1/4)E^((-1)^(3/8) a) - I E^(-(-1)^(5/8) a) 
- (-1)^(3/4)E^(-(-1)^(7/8) a))
Exp[I a/2 ( -a + Sqrt[4 + a^2]) t] Exp[-I  a x], {a, -Infinity, 0}]
 + NIntegrate[(E^(-(-1)^(1/8) a) + (-1)^(1/4)
E^(-(-1)^(3/8) a) - I E^((-1)^(5/8) a) - (-1)^(3/4)
 E^((-1)^(7/8) a)) Exp[-I a/ 2 ( a + Sqrt[4 + a^2]) t] Exp[-Ia x],
 {a, 0, Infinity}])], {x, -10, 10}, {t, 0, 5},  PlotPoints -> {40, 40}]
\end{verbatim}
}}

 \medskip
 
{\textbf{Observing velocity differences.}} 
Suppose that we have a sudden happening, and that various particles governed by the massless KG equation are emitted, some with low and some with high frequency. For $\omega$ negative and small (that is, compared to $|\lambda|^{-1}$) the difference between the low frequency speed $\cc$ and $\cc\,\sqrt{1-\mathrm{i}\,\lambda\,\omega}$ is small. Over a total time $T$ between creation and detection, the difference $\delta$ in the two signals arrival times (with larger $|\omega|$) arriving first) is 
\begin{eqnarray}
\delta\,=\,T(\sqrt{1-\mathrm{i}\,\lambda\,\omega}-1)\,\cong\, T\,|\lambda|\,|\omega|/2\ .
\end{eqnarray}
The higher frequency energy $E$ is approximately $\hbar\,|\omega|$, so we get
\begin{eqnarray}
E\,\cong\, \frac{2\,\delta\,\hbar}{T\,|\lambda|}\ .
\end{eqnarray}
To make $\delta$ as large as possible we might take $T$ to be very large -- $10^{10}$ years being about as old as we can observe signals. The time difference we can detect is more limited by the distant transmitter than our receivers, the problem is how synchronised were the different wavelength pulses originally. Let us be optimistic and guess 1 second for a compact source, as this is all at very best an order of magnitude argument. Then $\delta/T\cong 3\times 10^{-18}$. 
Now the energy for the $|\omega|$ large particles would be about
\begin{eqnarray}
E \,\cong\, \frac{6\times 10^{-52}}{|\lambda|}\,\cong\,10^{-8}\ \mathrm{Joules}\ ,
\end{eqnarray}
which is a lot for a small particle. It might be more familiar to quote it as approximately $10^{11}$ eV. This is 40 times less than the current single beam energy limit at CERN (two beams collide to increase the effective energy), and cosmic rays reach much higher energies. However this is deceptive, for two reasons. Firstly, charged particles have a problem with being scattered by magnetic fields.
Secondly, the sources of these very high energy cosmic rays (see e.g.\ \cite{CosmicRaySupernova}) are likely to be supernova remnants, and are not likely to show the very rapid variation that we have assumed (of the order of one second). However there are much more energetic cosmic rays, and in \cite{CosmicRayOrigin} an effective limit of about $5\times 10^{19}$ eV  is placed on cosmic ray particles from distant sources, due to interaction with the microwave background. This is about 500,000,000 times the energy we calculated above, and would correspond to a time delay of about 16 years in a journey of $10^{10}$ light years. We refer the reader to \cite{Am-SmoProspect,AmElMaNaTest,Am-CaMaj} for a more detailed account of possible measurements. The paper \cite{LimitOnVariationLight} refers to trying to observe a variation in light speed, but it is not obvious that the noncommutative geometry implies such a variation, as we explain in Section~\ref{chdsioaov}, with the large caveat that we do not discuss spinor fields. 

\section{Stars, conjugates and covariant derivatives} \label{hjdksvk}
The conjugate notation is not introduced for fun -- it is considerably more difficult to do differential geometry for noncommutative star algebras without it. 
Take $A$ to be an algebra, e.g.\ complex valued functions on space-time.
An $A$-bimodule $E$ contains things which can be multiplied by elements of $A$ on the left and right, obeying associativity rules for $a,b\in A$ and $e\in E$:
\begin{eqnarray}
a.(b.e)\,=\,(a\,b).e\ ,\ a.(e.b)\,=\,(a.e).b\ ,\ e.(a\,b)\,=\,(e.a).b\ .
\end{eqnarray}
Examples are forms or vector fields on manifolds, which can be multiplied by functions, but we have to take care on which side the multiplication happens.
The conjugate bimodule $\overline{E}$ contains elements $\overline{e}$ with products $\overline{e}.a=\overline{a^*.e}$ and 
$a.\overline{e}=\overline{e.a^*}$. Ignoring the distinction between $E$ and $\overline{E}$  would cause confusion.

It is usually convenient to describe a left covariant derivative on $E$ as a map $\nabla:E\to \Omega^1 A\tens_A E$ which obeys the left Liebniz rule, 
\begin{eqnarray}  \label{dghjkjgfx}
\nabla(a.e)\,=\,a.\nabla(e)+\extd a\tens e\ .
\end{eqnarray}
This can be converted to the covariant derivative in the direction of a vector field by the evaluation map from Section~\ref{cbvhdksukuv}. The symbol $\tens_A$ means that we take the fiberwise tensor product for vector bundles, or more technically that we set $\xi.a\tens e=\xi\tens a.e$. 

If $\nabla:E\to \Omega^1 A\tens_A E$ is a left covariant derivative, then $\overline{E}$ has a corresponding right covariant derivative $\nabla:\overline{E}\to \overline{E}\tens_A \Omega^1 A$ which obeys the right Liebniz rule, 
\begin{eqnarray}  \label{dghjkjgfxvdws}
\nabla(\overline{e}.a)\,=\,\nabla(\overline{e}).a+\overline{e}\tens\extd a\ .
\end{eqnarray}
This is given by the formula $\nabla(\overline{e})=\overline{k}\tens \eta^*$, where
$\nabla(e)=\eta\tens k$. More details can be found in \cite{BeMaStarRiem}.

The 1-forms $\Omega^1 A$ consist of sums of elements of the form $a.\extd b$ where $a,b\in A$. 
There is a star operation on $\Omega^1 A$ given by
$(a.\extd b)^*=\extd b^*.a^*$, and this gives a bimodule map $\star:\Omega^1A\to \overline{\Omega^1A}$ by $\star(\eta)=\overline{\eta^*}$. A 1-form $\xi$ is Hermitian if $\xi^*=\xi$. The star operation extends to 2-forms $\Omega^2 A$ by $(\xi\wedge\eta)^*=-\eta^*\wedge\xi^*$ for $\xi,\eta\in\Omega^1 A$.

\section{Electromagnetism, or $U_1$ gauge theory}  \label{vtyiuiuyf}
$U_1$ gauge theory is about covariant derivatives on sections of line bundles. As we are only dealing with the space-time $\mathbb{R}^4$ we shall assume that the bundle is trivial, which is a fancy way of saying that we just deal with derivatives of functions, i.e.\ we set $E=A$ in the language of Section~\ref{hjdksvk}. We shall only deal with Maxwell's theory, there will be no spinor fields. 

Classically the unitary functions are unit norm complex valued functions, and they form the local gauge transformation group. In noncommutative geometry we specify $u^*=u^{-1}$. 
Such unitary functions $u$ act on other functions by multiplication, and this action transfers to an action on covariant derivatives on functions. This becomes slightly more complicated in noncommutative geometry. 

Take a left covariant derivative $\nabla$ on $E=A$. This is uniquely specified by its value on the function 1, i.e.\ by $\xi\in\Omega^1A$ where $\nabla(1)=\xi\tens 1$. Likewise the curvature of $\nabla$ is given by the 2-form $R(1)=\extd\xi-\xi\wedge\xi$. On functions we set a function valued inner product by
\begin{eqnarray}
\<f,\overline{g}\>\,=\,f\,g^*\ .
\end{eqnarray}
If we have $\nabla(g)=\eta\tens k$, then $\nabla(\overline{g})=\overline{k}\tens \eta^*$ defines a right covariant derivative on $\overline{A}$. Having the covariant derivative preserve the Hermitian inner product is equivalent to $\xi^*=-\xi$. 

For the gauge group action, we might guess that a unitary $u\in A$ would act on a covariant derivative to give 
\begin{eqnarray}  \label{cvfjctyudxutxt}
(u\,\la\nabla)(e)\,=\,u\,\nabla(u^{-1}\,e)\ ,
\end{eqnarray}
but unfortunately $u\,\la\nabla$ does not obey the left Liebniz rule (\ref{dghjkjgfx}). Instead we use multiplication on the right, and define a right action by
\begin{eqnarray}  \label{cvfjctyudxutxttt}
(\nabla\ra\,u)(e)\,=\,\nabla(e\,u^{-1})\,u\ .
\end{eqnarray}
We consider $\xi\in\Omega^1A$ to be the gauge field, and by the left Liebniz rule we have a gauge action
\begin{eqnarray}
\xi\,\ra\,u\,=\, \extd u^*.u+u^*.\xi.u\ .
\end{eqnarray}
The gauge action on the curvature is $(R\,\ra\,u)(1)=u^*\,R(1)\,u$. The curvature contains the usual electric and magnetic fields.

Classically $\nabla$ has form $\nabla(1)=\mathrm{i}\,q\,A_a\extd x_a\tens 1$, and
we define $F_{ab}=\frac{\partial A_b}{\partial x_a}-\frac{\partial A_a}{\partial x_b}$. 
The
current corresponding to an electromagnetic field is real, and is given by the trace of the derivative
\begin{eqnarray}  \label{bchuadkskuyv}
\mathrm{j}^b\,=\, \partial_a\,F^{ab}
\end{eqnarray}
where $F^{ab}$ has with indices raised by the metric. For a given current, this is an equation of motion of the field. Finding a noncommutative analogue of (\ref{bchuadkskuyv}) is not entirely trivial, as we shall see.

\section{A trace of a derivative}
We need a version of (\ref{bchuadkskuyv}) that makes sense for noncommutative geometry. The guiding principle is that $F$ is derived from the curvature $R(1)$, which is an antiHermitian 2-form, and that $\mathrm{j}$ should be Hermitian, i.e.\ classically a real current. It is easier for us to consider everything to be a form, rather than vector fields, as there is a direct meaning of a form being Hermitian. Now we could always impose $\mathrm{j}$ being Hermitian, simply by giving some formula and averaging it with its star. However we would like a definition for the divergence which is natural as possible, given the Hermitian constraint. The key to this working is the choice of antisymmetry in the definition of the $F_{ab}$. Classically this is obvious, but it really depends on the fact that $\extd x_a$ and $\extd x_b$ anticommute. In fact, if this did not happen it would not be at all obvious what to do. In general, this would be related to the properties of a splitting map from 2-forms to the tensor product of two 1-forms (or equivalently, the interior product of a vector field with a 2-form). This is related to other things, such as a formula for the Ricci tensor, and is not well understood. 

Suppose that $\<,\>:\overline{\Omega^1 A}\tens_A\Omega^1 A\to A$ is a Hermitian form, i.e.\ it is an $A$-bimodule map and
\begin{eqnarray}
\<\overline{\xi},\eta\>^*\,=\, \<\overline{\eta},\xi\>\ .
\end{eqnarray}
We use the notation $ \langle \overline{dx_a}, dx_b\rangle = g^{ab} $.

\begin{defn} \label{105}
Define the following operations $ K,S:\overline{\Omega^1 A}\otimes_A\Omega^1 A\longrightarrow\Omega^1 A $ for a left covariant derivative $ \nabla $ on $ \Omega^1 A $:
\begin{eqnarray*}
K &=& (\langle,\rangle\otimes \id)(\nabla\otimes \id+\id\otimes\nabla)\ ,\\
S &=& (\id\otimes\langle,\rangle)(\star^{-1}\otimes\star\otimes \id)(\nabla\otimes \id+\id\otimes\nabla)\ .
\end{eqnarray*}
\end{defn}

%%%%%%%%%%%%%%
\begin{prop}\label{prp6} For $ \overline{\eta_1}\otimes\eta_2\in\overline{\Omega^1 A}\otimes_A\Omega^1 A $,
\begin{equation*}\label{107}
S(\overline{\eta_2}\otimes \eta_1)=K(\overline{\eta_1}\otimes\eta_2)^*
\end{equation*}
\end{prop}
\proof
Applying $ K $, where
$\nabla\eta_i=\zeta_i\tens\kappa_i$:
\begin{eqnarray*}
K(\overline{\eta_1}\otimes\eta_2) &=& \langle\overline{\zeta_1},\kappa_1^*\rangle\eta_2+\langle\overline{\eta_1},\kappa_2\rangle\zeta_2\ , \cr
K(\overline{\eta_1}\otimes\eta_2)^*&=&\eta_2^*\langle\overline{\kappa_1^*},\zeta_1\rangle+\zeta_2^*\langle\overline{\kappa_2},\eta_1\rangle\ .
\end{eqnarray*}
Now, applying $ S $ to $ \overline{\eta_2}\otimes\eta_1 $, we get:
\begin{equation*}
S(\overline{\eta_2}\otimes \eta_1)=K(\overline{\eta_1}\otimes\eta_2)^*\ .\qquad\blacksquare
\end{equation*}

%%%%%%%%%%%%%%
\begin{cor}\label{cr1}
If $ \eta_1^*\otimes\eta_2=\eta_2^*\otimes\eta_1 \in \Omega^1 A\otimes\Omega^1 A $, then $ (K-S)(\overline{\eta_1}\otimes\eta_2) $ is antiHermitian.
\end{cor}

Now we can try to make a noncommutative version of (\ref{bchuadkskuyv}). If we take a real multiple of $ R(1) $ to be $ \eta_1^*\wedge\eta_2 $, if $ \eta_1^*\otimes\eta_2 $ obeys the condition of Corollary~\ref{cr1} then $ (K-S)(\overline{\eta_1}\otimes\eta_2) $ is an antiHermitian trace of the derivative. 
We could use this to define a Hermitian current simply by multiplying by $\mathrm{i}$. Our problem is to satisfy the conditions of Corollary~\ref{cr1}, beginning with an antiHermitian 2-form.

\begin{prop}\label{prp7}
If $ m_{ab} \,\, dx_a\wedge dx_b $ is antiHermitian, then:
\begin{equation*}\label{110}
m_{ab} \,\, dx_a\wedge dx_b=\Big(m_{ba}^{\star}+\Lambda_{sp}^a \dfrac{\partial m_{bs}^{\star}}{\partial x_p}+\Lambda_{sp}^b \dfrac{\partial m_{sa}^{\star}}{\partial x_p}\Big)\, dx_a\wedge dx_b
\end{equation*}
\end{prop}
\proof
\begin{eqnarray*}
(m_{ab}\,\, dx_a\wedge dx_b)^{*} &=& -dx_b\wedge dx_a\,\, m_{ab}^{*}
\,=\, -dx_a\wedge dx_b\,\, m_{ba}^{*}\\
&=& -dx_a\wedge \,\, m_{ba}^{*}\,\,dx_b-dx_a\wedge \Lambda_{bp}^s\,\, \dfrac{\partial m^{*}_{ba}}{\partial x_p} dx_s\\
&=& -(m_{ba}^{*}+\Lambda_{sp}^a \dfrac{\partial m^{*}_{bs}}{\partial x_p}+\Lambda_{sp}^b \dfrac{\partial m^{*}_{sa}}{\partial x_p} )dx_a\wedge dx_b\ .\quad\blacksquare
\end{eqnarray*}

%%%%%%%%%%%%%%
\begin{cor}\label{cr2}
If $ m_{ab} $ is antisymmetric, i.e., $ m_{ab}=-m_{ba} $, and $ m_{ab}\,\, dx_a\wedge dx_b $ is antiHermitian, then:
\begin{equation*}\label{111}
m_{ab}=-m_{ab}^* +(\Lambda_{sp}^a \dfrac{\partial m_{bs}^{*}}{\partial x_p}+\Lambda_{sp}^b \dfrac{\partial m_{sa}^{*}}{\partial x_p})
\end{equation*}
\end{cor}

%%%%%%%%%%%%%%
\begin{cor}\label{cr3}
If $ m_{ab} $ is antisymmetric and $ m_{ab}\,\, dx_a\wedge dx_b $ is antiHermitian, then if $ \eta_1^* \otimes \eta_2= m_{ab}\,\, dx_a\otimes dx_b $, we have:
\begin{equation*}\label{112}
\eta_1^* \otimes \eta_2=\eta_2^* \otimes \eta_1
\end{equation*}
\end{cor}
\proof
Look at $ \eta_1^* \otimes \eta_2= m_{ab}\,\, dx_a\otimes dx_b $,
\begin{eqnarray*}
\eta_2^* \otimes \eta_1 &=& dx_b\otimes dx_a\,\, m_{ab}^*\\
&=& (m_{ab}^*+\Lambda_{sp}^a \dfrac{\partial m_{sb}^{*}}{\partial x_p}+\Lambda_{sp}^b \dfrac{\partial m_{as}^{*}}{\partial x_p}) dx_b\otimes  dx_a
\end{eqnarray*}
from (\ref{111}), this is
\begin{eqnarray*}
\eta_2^* \otimes \eta_1 &=& -m_{ab}\,\,dx_b\otimes  dx_a
\,=\, -m_{ba}\,\,dx_a\otimes  dx_b\\
&=& m_{ab}\,\,dx_a\otimes  dx_b
\,=\, \eta_1^* \otimes \eta_2\ .\quad\blacksquare
\end{eqnarray*}

%%%%%%%%%%%%%%
Now we have to calculate the antiHermitian form given by Corollary \ref{cr1} using the result of Corollary \ref{cr3}.

\begin{prop}\label{prp8}
If $ m_{ab} $ is antisymmetric and $ m_{ab}\, dx_a\wedge dx_b $ is antiHermitian, then $ (K-S)(\overline{(m_{ab}\, dx_a)^*}\otimes dx_b) $ is antiHermitian, and
\begin{eqnarray*}\label{113}
:\!(K-S)(\overline{(m_{ab}\, dx_a)^*}\otimes dx_b)\!: \,=\, g^{aq}\left(\begin{array}{c}:\!\dfrac{\partial m_{ab}^*}{\partial x_q}\!:-:\!\dfrac{\partial m_{ab}^*}{\partial x_q}\!:^* -\lambda\sum x_i \dfrac{\partial^3 :\!m_{ab}\!:^*}{\partial x_0\partial x_i\partial x_q}\\ -\Lambda_{sp}^a \dfrac{\partial^2 (:\!m_{sb}\!:^*+:\!m_{sb}\!:)}{\partial x_p\partial x_q} +\Lambda_{rn}^b\dfrac{\partial^2 :\!m_{ar}\!:}{\partial x_n\partial x_q} \end{array}\right)dx_b
\end{eqnarray*}
\end{prop}
%%%%%%%%%%%%%%
\proof
Substituting for $ dm^*_{ab} $, and using $ \langle \overline{dx_a}, dx_b\rangle = g^{ab} $ gives:
\begin{equation*}
\langle\overline{\zeta_1},\kappa_1^*\rangle\eta_2=(g^{aq}\dfrac{\partial m_{ab}^*}{\partial x_q}-\Lambda_{sp}^a\dfrac{\partial^2 m_{sb}^*}{\partial x_p \partial x_q}g^{aq})dx_b
\end{equation*}
remembering that the partial derivative of an algebra element $ m^*_{ab} $ is defined in Section~\ref{cbvhdksukuv}. We also need:
\begin{eqnarray*}
(\langle\overline{\zeta_1},\kappa_1^*\rangle\eta_2)^* &=& g^{aq}dx_b \big( (\dfrac{\partial m_{ab}^*}{\partial x_q})^*+\Lambda_{sp}^a(\dfrac{\partial^2 m_{sb}^*}{\partial x_p \partial x_q})^*\big)\\
&=& g^{aq}(\dfrac{\partial m_{ab}^*}{\partial x_q})^* dx_b+g^{aq} \Lambda_{sp}^a(\dfrac{\partial^2 m_{sb}^*}{\partial x_p \partial x_q})^* dx_b - g^{aq} \Lambda_{bn}^r\dfrac{\partial}{\partial x_n}(\dfrac{\partial m_{ab}^*}{\partial x_q})^* dx_r\ ,
\end{eqnarray*}
so
\begin{eqnarray*}
\langle\overline{\zeta_1},\kappa_1^*\rangle\eta_2 - (\langle\overline{\zeta_1},\kappa_1^*\rangle\eta_2)^* &=& g^{aq}\left(\begin{array}{c}\dfrac{\partial m_{ab}^*}{\partial x_q}-\Lambda_{sp}^a\dfrac{\partial^2 m_{sb}^*}{\partial x_p \partial x_q}-(\dfrac{\partial m_{ab}^*}{\partial x_q})^*\\ -\Lambda_{sp}^a(\dfrac{\partial^2 m_{sb}^*}{\partial x_p \partial x_q})^*)
+\Lambda_{rn}^b\dfrac{\partial}{\partial x_n}(\dfrac{\partial m_{ar}^*}{\partial x_q})^* \end{array}\right) dx_b
\end{eqnarray*}
If we take the normal order of this we get, as the star and partial derivatives are the usual ones to $ \lambda^0 $, to $ O(\lambda) $
\begin{eqnarray*}
\langle\overline{\zeta_1},\kappa_1^*\rangle\eta_2 - (\langle\overline{\zeta_1},\kappa_1^*\rangle\eta_2)^* &=& g^{aq}\left(\begin{array}{c}:\!\dfrac{\partial m_{ab}^*}{\partial x_q}\!:-:\!(\dfrac{\partial m_{ab}^*}{\partial x_q})^*\!:-\Lambda_{sp}^a\dfrac{\partial^2 :\!m_{sb}\!:^*}{\partial x_p \partial x_q}\\ -\Lambda_{sp}^a \dfrac{\partial^2 :\!m_{sb}\!:}{\partial x_p \partial x_q}
+\Lambda_{rn}^b\dfrac{\partial^2 :\!m_{ar}\!:}{\partial x_n \partial x_q} \end{array}\right) dx_b
\end{eqnarray*}
Now, using (\ref{bvhislzggi}), we get the answer.\quad$\blacksquare$

%%%%%%%%%%%%%%
\begin{cor}\label{cr4}
If $ m_{ab} $ is antisymmetric and $ m_{ab}dx_a\wedge dx_b $ is antiHermitian, then $ (K-S)(\overline{(m_{ab}\, dx_a)^*}\otimes dx_b) $ is antiHermitian 1-form, and using the 1-1 correspondence in Proposition~\ref{bvcahjksjhc} it corresponds to a classical antiHermitian 1-form $ g^{aq}(\dfrac{\partial :\!m_{ab}\!:^*}{\partial x_q}-\dfrac{\partial :\!m_{ab}\!:}{\partial x_q})dx_b $ 
\end{cor}
\proof
To zeroth order in $ \lambda $, the $ m_{ab} $ are imaginary. Now, remembering that $\lambda$ is imaginary, the other antiHermitian parts are $O(\lambda^2)$.\quad$\blacksquare$

\section{The current}
The curvature $R(1)$ is the difference of the following two terms:
\begin{eqnarray}
\qquad:\!\extd\xi\!: &=&  \frac{\partial :\!\xi_b\!:}{\partial x_a}\, \extd x_a\wedge\extd x_b - \lambda\,\Big(
\frac{\cc^2}{2}\, \frac{\partial^2 \!  :\!\xi_b\!:}{\partial x_i^2} \, \extd x_0
+ \frac{\partial^2  :\!\xi_b\!:}{\partial x_0\,\partial x_i} \, \extd x_i
\Big)\wedge\extd x_b\ ,\cr
:\!\xi\wedge\xi\!: &=& \lambda\,\Big(x_i\,\frac{\partial :\!\xi_a\!:}{\partial x_i}\,\frac{\partial :\!\xi_b\!:}{\partial x_0}\,\extd x_a
- :\!\xi_0\!: \frac{\partial :\!\xi_b\!:}{\partial x_i}\,\extd x_i \cr
&& -\  \cc^2\, :\!\xi_i\!:\, \frac{\partial :\!\xi_b\!:}{\partial x_i}\,\extd x_0
\Big)\wedge\extd x_b\ .
\end{eqnarray}
Note that the second term is nonzero purely due to noncommutative effects. 
Then $ R(1)=R(1)_{ab}\, dx_a\wedge dx_b $, where:
\begin{eqnarray}\label{121}
:\!R(1)_{ab}\!: &=& \dfrac{\partial:\!\xi_b\!:}{\partial x_a} - \dfrac{\lambda\, c^2}{2}\,\dfrac{\partial^2:\!\xi_b\!:}{\partial x^2_i}\, \delta_{a,0} - \lambda\,\dfrac{\partial^2:\!\xi_b\!:}{\partial x_0 \partial x_i}\, \delta_{a,i}\nonumber \\
&&- \lambda\,\dfrac{\partial:\!\xi_a\!:}{\partial x_i}\, x_i\,\dfrac{\partial:\!\xi_b\!:}{\partial x_0} + \lambda\, c^2\, :\!\xi_i\!:\,\dfrac{\partial:\!\xi_b\!:}{\partial x_i}\,\delta_{a,0}\nonumber \\ 
&&+ \lambda\, :\! \xi_0\!:\,\dfrac{\partial:\!\xi_b\!:}{\partial x_i}\,\delta_{a,i}\ .
\end{eqnarray}
To ensure antisymmetry, define $q\,m_{ab}=R(1)_{ab}-R(1)_{ba}$. 
Then $ q\,m_{ab} dx_a\wedge dx_b=2R(1) $, as $ dx_a\wedge dx_b=-dx_b\wedge dx_a $. Also $ R(1) $ is antiHermitian as $ \xi $ is antiHermitian. 
Remembering that $ \lambda $ is imaginary, as $:\!\xi_a\!:$ is antiHermitian to zeroth order, the second and third terms of 
(\ref{121}) are Hermitian to $O(\lambda^2)$. If we put the antiHermitian part of $:\!\xi_a\!:$ equal to $\mathrm{i}\,q\,A_a$, then
\begin{eqnarray}\label{125}
\text{AntiHermitianPart}(:\!m_{ab}\!:) &=& \mathrm{i}\, (\dfrac{\partial A_b}{\partial x_a}-\dfrac{\partial A_a}{\partial x_b})\nonumber\\
&&+ \lambda\,q\, (\dfrac{\partial A_a}{\partial x_i}\,x_i\,\dfrac{\partial A_b}{\partial x_0}-\dfrac{\partial A_b}{\partial x_i}\,x_i\,\dfrac{\partial A_a}{\partial x_0})\nonumber\\
&&- \lambda\,q\,c^2\,A_i(\dfrac{\partial A_b}{\partial x_i}\, \delta_{a,0}-\dfrac{\partial A_a}{\partial x_i}\, \delta_{b,0})\nonumber\\
&&- \lambda\,q\, A_0(\dfrac{\partial A_b}{\partial x_i}\,\delta_{a,i}-\dfrac{\partial A_a}{\partial x_i}\,\delta_{b,i})\ .
\end{eqnarray}
By Corollary~\ref{cr4} the following formula (\ref{149}) defines a Hermitian 1-form $\mathrm{j}$:
\begin{equation}\label{149}
-\,2\,\mathrm{i\,j}=(K-S)(\overline{(m_{ab}\, dx_a)^*}\otimes dx_b)
\end{equation}
Further by Corollary~\ref{cr4}, this corresponds to the following formula (\ref{150}) for the Hermitian part of the normal order of  $\mathrm{j}$:
\begin{equation}\label{150}
-\,2\,\mathrm{i} \  \mathrm{HermitianPart}(:\!\mathrm{j}\!: ) \,=\,
g^{aq}\,\dfrac{\partial(:\!m_{ab}\!:^*-:\!m_{ab}\!:)}{\partial x_q}\,dx_b
\end{equation}
%%%%%%%%%%%%%%
Substituting (\ref{125}) into (\ref{150}) gives
\begin{eqnarray}\label{128}
\mathrm{HermitianPart}(:\!\mathrm{j}\!: ) &=& g^{aq}\dfrac{\partial}{\partial x_q}\left(\begin{array}{c}(\dfrac{\partial A_b}{\partial x_a}-\dfrac{\partial A_a}{\partial x_b})\nonumber\\
- i\,\lambda\,q\, x_i(\dfrac{\partial A_a}{\partial x_i}\,\dfrac{\partial A_b}{\partial x_0}-\dfrac{\partial A_b}{\partial x_i}\,\dfrac{\partial A_a}{\partial x_0})\nonumber\\
+ i\,\lambda\,q\,c^2\,A_i(\dfrac{\partial A_b}{\partial x_i}\, \delta_{a,0}-\dfrac{\partial A_a}{\partial x_i}\, \delta_{b,0})\nonumber\\
+ i\,\lambda\,q\, A_0(\dfrac{\partial A_b}{\partial x_i}\,\delta_{a,i}-\dfrac{\partial A_a}{\partial x_i}\,\delta_{b,i})\end{array}\right)dx_b\ .
\end{eqnarray}
This is simply the usual equation for the current;
\begin{eqnarray}\label{151}
\mathrm{HermitianPart}(:\!\mathrm{j}\!: )
&=& g^{aq}\,\dfrac{\partial}{\partial x_q}\, (F_{ab})\, dx_b
\end{eqnarray}
but for a modified field;
\begin{eqnarray}\label{152}
F_{ab} &=& \dfrac{\partial A_b}{\partial x_a}-\dfrac{\partial A_a}{\partial x_b} - \mathrm{i}\,\lambda\, q\, x_i(\dfrac{\partial A_a}{\partial x_i}\,\dfrac{\partial A_b}{\partial x_0}-\dfrac{\partial A_b}{\partial x_i}\,\dfrac{\partial A_a}{\partial x_0})\nonumber\\
&&+\, \mathrm{i}\,\lambda\,c^2\,q\,A_i(\dfrac{\partial A_b}{\partial x_i}\, \delta_{a,0}-\dfrac{\partial A_a}{\partial x_i}\, \delta_{b,0}) + \mathrm{i}\,\lambda\,q\, A_0(\dfrac{\partial A_b}{\partial x_i}\,\delta_{a,i}-\dfrac{\partial A_a}{\partial x_i}\,\delta_{b,i})\ .
\end{eqnarray}
Note that \cite{HaJuMeElectro} also gives nonlinear terms in electromagnetism. 
The presence of an explicit $x_i$ in the formula (\ref{152}) for $F_{ab}$ may seem alarming, as after all its entries are electric and magnetic fields, and therefore possibly measurable. However this term can be considered an artifact of the normal ordering, as from (\ref{vcdjkbvhjk})
\begin{eqnarray} \label{vcdjkbcdsvhjk}
:\![\psi,\chi]\!: \,=\, \lambda\sum_i x_i\,\Big(\frac{\partial :\!\psi\!:}{\partial x_i}\, \frac{\partial :\!\chi\!:}{\partial x_0}
- \frac{\partial :\!\chi\!:}{\partial x_i}\, \frac{\partial :\!\psi\!:}{\partial x_0}\Big)\ .
\end{eqnarray}

\section{Electromagnetic plane waves} \label{chdsioaov}
Set $A_a=E_a\,\sin(\alpha_0\,t+\alpha_1\,x_1+\beta_a)$ with $\beta_0=0$. Then (\ref{152}) gives the only nonzero $F_{ab}$ as the following, up to antisymmetry:
\begin{eqnarray}\label{1525}
F_{01} &=& \dfrac{\partial A_1}{\partial x_0}-
(1+\mathrm{i}\,\lambda\,q\, A_0)\,\dfrac{\partial A_0}{\partial x_1} + \mathrm{i}\,\lambda\,q\,c^2\,A_1\,\dfrac{\partial A_1}{\partial x_1}   \ ,\cr
F_{0j} &=& \dfrac{\partial A_j}{\partial x_0}+ \mathrm{i}\,\lambda\,c^2\,q\,A_1\,\dfrac{\partial A_j}{\partial x_1}   \ ,\quad j\neq 0,1\ ,\cr
F_{1j} &=&  (1+\mathrm{i}\,\lambda\,q\, A_0)\,\dfrac{\partial A_j}{\partial x_1}\ ,\quad j\neq 0,1\ .
\end{eqnarray}
For zero current we get
\begin{eqnarray}\label{1515}
0 &=& g^{00}\,\dfrac{\partial \,F_{0b}}{\partial x_0} + 
g^{11}\,\dfrac{\partial \,F_{1b}}{\partial x_1}\ .
\end{eqnarray}
From this we deduce that $F_{01}$ is constant, so 
\begin{eqnarray}\label{152cwds5}
A_0\,\dfrac{\partial A_0}{\partial x_1} &=& c^2\,A_1\,\dfrac{\partial A_1}{\partial x_1}   \ ,\cr
E_0^2\,\sin(2(\alpha_0\,t+\alpha_1\,x_1)) &=& c^2\,
E_1^2\,\sin(2(\alpha_0\,t+\alpha_1\,x_1+\beta_1))\ .
\end{eqnarray}
We deduce $\beta_1=0$ or $\beta_1=\pi$ and $E_0^2=c^2\,
E_1^2$. We also deduce that $\alpha_0\,E_1=\pm\,\alpha_1\,E_0$. 
Unless $E_0=E_1=0$ we must have $\alpha_0=\pm\,\cc\,\alpha_1$, so the plane wave travels at speed $\cc$. However if $E_0=E_1=0$ then (\ref{1525}) give a special case of the classical equations, so the speed is also $\cc$. We deduce that all zero current electromagnetic plane waves travel at speed $\cc$. However it would be dangerous to extrapolate to saying that all electromagnetic plane waves \textit{in vacuum} travel at speed $\cc$. It would not be impossible for some mechanism to induce a current of size about $|\lambda|$ in a vacuum, in fact such an effect might be likely, as the following hand waving argument shows. Fields acting on spontaneously produced charged particle pairs would accelerate them in different directions. If the field reverses (as it would in a traveling wave), they would be accelerated back together, to recombine. This would constitute a current which would likely increase as the frequency increases, as the expected time to recombination would be reduced.

\section{Quantum theory, gravity and noncommutative geometry}
Science is about observation, experiment and prediction. To observe requires measurement, and that is limited by our knowledge and technology. To teach the theories of quantum mechanics three thousand years ago would have been, from an experimental point of view, completely pointless. The technology to make observations of the theory simply did not exist. To advance, we need to base our observations of possible new phenomena on our current understanding. For proposed physical applications of noncommutative geometry, we need to give evidence that we can in principle measure. This is what was so interesting about wave velocities in \cite{Am-CaMaj}. 

However, more generally, there are still problems in directly relating noncommutative geometry to concrete physical problems. 
There are some notable contributions: 
In \cite{DoFrRoQuantStruct,DoMoPiHorizon} the measurement problem of determining position very accurately is discussed, with the limiting phenomenon being the formation of trapped surfaces at very high energy densities, including detailed uncertainty relations.
The problem of geodesics related to the generalised uncertainty principle is discussed in \cite{ghoshQGrGeo}. 
We give four thought experiments to think about, and would be happy to hear about any previous discussion of related matters in the literature.

\begin{example}
Forget quantum gravity, and take the (in normal circumstances) much larger effects of quantum theory on gravity. As an example, an asteroid space ship is being steered by an autopilot. The autopilot observes a radioactive sample for a time. If there is a decay, it steers in one direction, if not it steers in another. The quantum uncertainty has been magnified into an observable gravitational effect. We might describe the result of a quantum experiment by a wave function. But how do we mathematically describe the action of this wave function on Riemannian geometry? (Please do not suggest taking an expectation of the stress-energy tensor, as the asteroid space ship has a chance of being in two possible places once a measurement is taken, and no chance of being half way between them or splitting into two smaller asteroids.)
Does this mean that, from one point of view, geometry has become `wave function valued'? A complete description of the situation should include a method of calculating with positions and momenta of test particles under the gravitational influence of the asteroid. Lest it be suggested that a multi-particle Schr\"odinger equation with gravitational potential would suffice, we might also let the asteriod be gravitationally towing a small black hole with orbiting test particles...
\end{example}

\begin{example} \label{gccddilskuyf}
Consider a lattice of widely spaced spheres in a very large room. Each sphere contains an atomic clock and a laser transmitter and receiver. Each has a recorder to write the times of sending and receiving laser pulses. Each has its position in the lattice written on its recorder. Now the clocks are synchronised, and the walls of the room are removed, revealing that the lattice is in deep space, in uniform motion towards the vicinity of a distant neutron star. The probes travel on, repeatedly transmitting laser pulses between themselves, and recording the data. After many years they are collected, and the data analysed. 

A classic dilemma ensues: To use low frequency pulses with low momentum but low spatial resolution, or high frequency pulses with high spatial resolution but high momentum. To this we now add the gravitational effect of the pulses themselves. Perhaps it might be possible to use the symmetry of the lattice to get these complicating effects to largely cancel out. However will such techniques survive the loss of symmetry imposed by the gravitational field of the neutron star? As the purpose of the spheres is to measure the geometry, we cannot argue that knowing the geometry allows us to devise a cancellation technique.

The spatial coordinates are quite simple - they are written on the recorders. Presumably we have a rather discretised version of $\mathbb{R}^3$, with commuting coordinates. However the time is less certain. But what stands out as taking the full force of the quantum uncertainty is the metric tensor, which is encoded in the measurements of the distances between the spheres. Does the uncertainty in the geometry manifest itself as noncommutative geometry in a calculable way? 
\end{example}

\begin{example}
People have considered the effect of noncommutative geometry on the anisotropy of the cosmic microwave background, e.g.\ 
\cite{akBaSaQu}. However is there a possibility that the CMB might give rise to noncommutativity? There are many possible noncommutative geometries, so what might `choose' between them a `physical' choice? In the first approximation, it should be a deformation of Minkowski space and its Lorentz symmetry. How to pick out natural deformations with a physical origin? Of course, such a deformation might be a product of some symmetry breaking, and be essentially `random', or even different in different places. However the dominant feature breaking Lorentz symmetry on a large scale in our universe is the CMB. Is there any plausible mechanism whereby the CMB could induce a noncommutativity in the geometry? There is at least one way in which the CMB interacts with quantum measurement processes, and therefore possibly with geometry: Very high energy photons can interact with the CMB photons, and may produce electron-positron pairs (see e.g.\ \cite{ChengCheng}).  In the classic Heisenberg uncertainty argument about simultaneous measurement of momentum and position, there is an implicit idea that the position could be determined with arbitrary precision by using higher energy photons. In the presence of the CMB there is a definite physical effect influencing these higher energy photons, and impacting any thought experiment on quantum measurement of distance. Similarly to Example~\ref{gccddilskuyf} (where we discussed gravitational effects) the problem is whether the effects on quantum measurement produce a geometry which effectively seems noncommutative.  Of course, any local effects of the CMB would easily be swamped by a nearby star, but in the vast darkness of intergalactic space it may be a dominant effect. 
\end{example}

\begin{example}
As a very specific example, just suppose that the noncommutative Schr\"odinger equation
of Section~\ref{cvgfkucjy} was backed by experimental evidence and generally accepted. The natural and useful thing to do would be to apply the Schr\"odinger equation to several particles. But now we have a problem: If we have vector coordinates $x$ and $x'$ for the two particles, what are the commutation relations between the various components of $x$ and $x'$? 
Would the two copies of spatial $\mathbb{R}^3$ commute with each other?  As the single particle noncommutative Schr\"odinger equation seems to incorporate some form of gravitational self-interaction, it seems quite likely that they would not.
We have little idea how to consider multiple copies of space-time - possibly some idea of braiding would be important here, but how would we cope with having a common time for both? 
\end{example}


\begin{thebibliography}{ggghhh}



\bibitem{LimitOnVariationLight}
Abdo A.A.\ et al., A limit on the variation of the speed of light arising from quantum gravity effects, 
Nature 462, 331-334, 2009

\bibitem{CosmicRaySupernova}
Ackermann M.\ et al., Detection of the Characteristic Pion-Decay Signature in Supernova Remnants,  Science
Vol. 339 no.\ 6121 pp 807-811 (2013)    
%  DOI: 10.1126/science.1231160

\bibitem{akBaSaQu}
Akofor E., Balachandran A.P., Jo S.G., Joseph A.\ \& Qureshi B.A.,
Direction-Dependent CMB Power Spectrum and Statistical Anisotropy from Noncommutative Geometry, JHEP 0805 (2008)

\bibitem{Am-SmoProspect}
Amelino-Camelia G.\ \& Smolin L., Prospects for constraining quantum gravity
dispersion with near term observations.  Phys.\ Rev.\ D 80:084017, 2009

\bibitem{AmElMaNaTest}
Amelino-Camelia G., Ellis J., Mavromatos N.E., Nanopoulos, D.V.\ \& Sarkar S.,
Tests of quantum gravity from observations of gamma-ray bursts. Nature 393,
763Ð765 (1998).

\bibitem{Am-CaMaj}
Amelino-Camelia G.\ \& Majid S.,
Waves on noncommutative space-time and gamma ray bursts,
Int.\ J.\ Mod.\ Phys.\ A 15, 4301 (2000).


\bibitem{BeMaStarGravInd}
 Beggs E.J.\ \& Majid S., `Gravity induced from quantum spacetime', arXiv:1305.2403
 
  \bibitem{barcats}
 Beggs E.J.\ \& Majid S., Bar categories and star operations, 
{\it Algebras and Representation Theory,} {12} (2009) 103-152.


\bibitem{BeMaStarRiem}
 Beggs E.J.\ \& Majid S.,`*-compatible connections in noncommutative Riemannian geometry', Journal of Geometry and Physics 61 (2011)  95-124.
 
  
 \bibitem{BeMaSemiClass}
 Beggs E.J.\ \& Majid S.,  `Semi-classical differential structures', Pacific Journal of Mathematics, Vol. 224 No. 1 (2006), p 1-44
 
 \bibitem{BeMaNonAssTwist}
 Beggs E.J.\ \& Majid S., `Nonassociative Riemannian geometry by twisting', in Quantum - a Festschrift for Tony Sudbery, Journal of Physics: Conference Series 254, 2010.
 

 \bibitem{berrySuper}
Berry M.V., Superluminal speeds for relativistic random waves,
    Journal of Physics A: Math.\ and Theor., 
    Vol.\ 45 no.\ 18, 

\bibitem{CosmicRayOrigin}
Biermann P.L., The origin of the highest energy cosmic rays, J.\ Phys.\ G: Nucl.\ Part.\ Phys. 23 no. 1
(1997)
%   doi:10.1088/0954-3899/23/1/002

\bibitem{BHMnonass}
{Bouwknegt\, P., Hannabuss\, K.C., Mathai\, V.,}
{Nonassociative tori and applications to T-duality, Commun.\ Math.\ Phys.\
264 (2006) 41-69.}

\bibitem{ChengCheng}
Cheng L.X.\ \&  Cheng K.S., Delayed MeVGeV Gamma-Ray Photons in Gamma-Ray Bursts: An Effect of Electromagnetic Cascades of Very High Energy Gamma Rays in the Infrared/Microwave Background, The Astrophysical Journal, 459:L79L82, 1996

\bibitem{DiJoMo} 
Dimitrijevi\'c M., Meyer F., M\"oller L.\ \& Wess J., Gauge theories on the kappa-Minkowski
spacetime, Eur. Phys. J. C 36 (2004) 117

\bibitem{DiJo} 
Dimitrijevi\'c M.\ \& Jonke L.,
A twisted look on kappa-Minkowski: $U(1)$ gauge theory, Journal of High Energy Physics, 2011:80, 2011

\bibitem{DoPe} 
Dobrev V.K.\ \& Petrov S.T.,
Q-Plane Wave Solutions of Q-Maxwell Equations, Quantum theory and symmetries, Proc.\ Second International Symposium 2001 in Krak\'ow. Eds.\ Edward Kapuscik \& Andrzej Horzela. World Sci., 2002.

\bibitem{DoFrRoQuantStruct}
Doplicher S., Fredenhagen K.\ \& Roberts J.E.,
The Quantum Structure of Spacetime at the Planck
Scale and Quantum Fields, Commun.\ Math.\ Phys.\ 172, 187-220 (1995)

\bibitem{DoMoPiHorizon}
Doplicher S., Morsella G.\ \& Pinamonti N.,
On Quantum Spacetime and the horizon problem, arXiv:1201.2519

\bibitem{DuSi} 
Durhuus B.\ \& Sitarz A.,
Star product realizations of kappa-Minkowski space, arXiv:1104.0206

\bibitem{ghoshQGrGeo}
Ghosh S., Quantum Gravity Corrected Geodesic Motion
and Violations of Equivalence Principle, arXiv:1303.1256

\bibitem{ghoshPalKappa}
Ghosh S.\ \& Pal P., $\kappa$-Minkowski spacetime through exotic ``oscillator", Phys.\ Lett.\ B
Vol.\ 618, Issues 1-4,  2005, p 243-251.

\bibitem{HarikMaxwells}
Harikumar E., Maxwell's equations on the $\kappa$-Minkowski spacetime and Electric-Magnetic duality, Europhys.\ Lett.\ 90 (2010) 21001

\bibitem{HaJuMeElectro}
Harikumar E., Juri\'c T.\ \& S. Meljanac S., Electrodynamics on $\kappa$-Minkowski space-time, Phys. Rev. D 84 (2011)

\bibitem{hawkRig} Hawkins, E.,  Noncommutative rigidity, Comm.\ Math.\ Phys., 246:211-235, 2004. 
 
\bibitem{BKellerAinfty}
Keller B., Introduction to {$A$}-infinity algebras and modules, Homology Homotopy Appl. Volume 3, Number 1 (2001), 1-35.

\bibitem{KonSoi09}
Kontsevich M.\ \& Soibelman Y., Notes on $A_\infty$-Algebras, $A_\infty$-Categories and Non-Commutative Geometry, 
Homological Mirror Symmetry,
Lecture Notes in Physics Volume 757, 2009, pp 1-67, Springer

\bibitem{majid12}
{Majid S.,} {\sl Almost commutative Riemannian geometry: wave operators, Commun.\ Math.\ Phys.\ 310, 569-609 (2012)}

\bibitem{majidNonAss}
Majid S., Gauge theory on nonassociative spaces, J. Math. Phys. 46 (2005) 103519.

\bibitem{majidNewtonianGrav}
Majid S., Newtonian gravity on quantum spacetime,  in press Euro Phys. J. Web of Conferences,  2013

\bibitem{majid-ruegg}
{Majid S.\ \& Ruegg H.,} {\sl Bicrossproduct structure of $\kappa$-Poincar\'e group and non-commutative geometry, Physics Letters B,
Volume 334, Issues 3-4, August 1994, P 348-354}

\bibitem{MercatiKappa}
Mercati F., Quantum kappa-deformed differential geometry and field theory, ArXiv:1112.2426.

\bibitem{PrGhPa}
Pramanik S., Ghosh S.\ \&, Pal P.,
Planck Scale Effects in Electrodynamics of a Generalized Charged Particle, arXiv:1212.6881


\bibitem{ryderbook}
Ryder L.H., Quantum field theory, 2nd.\ ed., C.U.P.\ 1996.


\bibitem{ShoreOptics}
Shore G.M., Quantum gravitational optics, Contemporary Physics, Volume 44, Issue 6, 2003

\bibitem{StasheffPentagon}
 Stasheff J.D., Homotopy associativity of H-spaces I \& II, Trans.\ Amer.\ Math.\
Soc.\ 108 (1963), 275-292, 293-312.

\bibitem{woron87}
{Woronowicz S. L.,} {\sl Twisted ${\rm SU}(2)$ group. An example of a noncommutative differential calculus.
Publ. Res. Inst. Math. Sci. 23 (1987), no. 1, 117--181.}

\end{thebibliography}
\end{document}